# Dynamics of entanglement in a one-dimensional Ising chain

*G.B. Furman[1,2], V. M. Meerovich[1], and V.L. Sokolovsky[1]*

[1]Physics Department, Ben Gurion University, Beer Sheva, POB 653, 84105, Israel

[2]Ohalo College, Qazrin, POB 222, 12900, Israel

Abstract:

The evolution of entanglement in a one-dimensional Ising chain is numerically studied under various initial conditions. We analyze two problems concerning the dynamics of the entanglement: (i) generation of the entanglement from the pseudopure separable state and (ii) transportation of the entanglement from one end of the chain to the other. The investigated model is a one-dimensional Ising spin-1/2 chain with nearest-neighbor interactions placed in an external magnetic field and irradiated by a weak resonant transverse field. The possibility of selective initialization of partially entangled states is considered. It was shown that, in spite of the use of a model with the direct interactions between the nearest neighbors, the entanglement between remote spins is generated.

## I. Introduction

The important role of entanglement [1] as a fundamental resource for quantum information process that oversteps the classical limits is particularly obvious and has been experimentally verified [2-6]. Direct interactions between particles are used as typical conditions to create entanglement [2-10]. However, in most systems with short-range interactions, entanglement between a pair of particles decays rapidly with the distance [4-11]. A much desired goal would be the ability to create entanglement between distant quantum objects which are not connected by direct interactions.

The static properties of the entanglement in spin chains has been extensively studied [1, 11, 12]. Recently, the study was extended into the understanding of entanglement dynamics in the one-dimensional *XX*, *XY* and Ising model systems [6, 12 -17]. Because a general theory of multipartite entanglement is not completed [1, 6, 11], many investigations were restricted by the study of dynamics of nearest neighbor qubit pairs [10, 16]. To understand the appearance and transport of entanglement of the system with number of qubits, it is of great interest to study the entanglement dynamics between pairs of remote qubits which are not connected by direct interactions.

Recently, it has been demonstrated that, for both an idealized one-dimensional Ising spin-1/2 chain with nearest-neighbor interactions [18] and a realistic spin-1/2 chain including the natural dipole-dipole interactions [19], placed in an external magnetic field and irradiated by a weak resonant transverse field, a wave of flipped spins can be triggered by a single spin flip. The qualitative explanation of this phenomenon is the following. When a spin it has two neighbors in the same state, the interaction with the neighbor(s) makes the spin off-resonant and the irradiation field does not change its state. The same happens when the spin is at the either end of the chain. When the two neighbors of a spin are in different states, the resonant irradiation field flips the spin. Therefore, if all spins are in the same state, the state of the entire system is stationary. If the first spin is flipped, its neighbor becomes resonant and flips, and then the next spin flips, and so on, resulting in macroscopic changes of the spin system, so called "quantum domino dynamics". Recently, the "quantum domino dynamics" was realized in the experiment [20]. The equivalent description of the operation can be given using a sequence of quantum logic gates [18, 21]

$$U = CNOT_{N-1,N} CNOT_{N-2,N-1} ... CNOT_{1,2} , \qquad (1)$$

which is a chain of unitary controlled-not gates $CNOT_{m,n}$. The $CNOT_{m,n}$ gate flips the target qubit $n$ when the control qubit $m$ is in the state $|1\rangle_m$ and does not change the qubit $n$ when the qubit $m$ is in the state $|0\rangle_m$. If the $1-th$ qubit is in the state $|1\rangle_1$, it flips qubit $2$, then qubit $2$ flips qubit $3$, and so on. The most important feature of the spin dynamics described above is signal amplification, when a state of the polarization of a single spin is converted into the total polarization of the spin system. Change in the total polarization (magnetization) of a spin system quantifies the efficiency of this system as a quantum amplifier.

In this paper we focus on the study of entanglement dynamics in a one-dimensional finite Ising chain of nuclear spins of 1/2. Numerical solutions for the concurrence dynamics are obtained for linear chains consisting up to seven spins and with various initial conditions.

We concentrate our investigation on two problems. The first one is the generation of the entanglement between two spins starting from the disentangled state. The second problem is the transportation of the entanglement of initially entangled spins along a spin chain. Thus, we obtain important information characterizing the entanglement between every two spins in the linear chain, including the spins which not interact directly. The influence of nearest and remote spins to the bipartite entanglement is studied.

## II. Hamiltonian of the model

Let us consider a one-dimensional (1D) Ising chain with nearest-neighbor interactions, irradiated by a weak transverse field of the resonance frequency [18]. The spin dynamics is described by the Liouville–von Neumann equation [22, 23]

$$i\frac{d\rho(t)}{dt} = [H(t), \rho(t)] \qquad (2)$$

with the Hamiltonian

$$H(t) = \sum_{n=1}^{N} \{\omega_0 I_n^z + 2\omega_1 I_n^x \cos\omega_0 t\} + J \sum_{n=1}^{N-1} I_n^z I_{n+1}^z \qquad (3)$$

where $\omega_0$ is the energy difference between the excited and ground states of an isolated spin, $J$ is the coupling constant, $\omega_1$ is the amplitude of irradiation field, $I_n^x$ and $I_n^z$ are the projections of the angular spin momentum operators on the axes $x$ and $z$, respectively, $N$ is the total number of spins in the chain. Spin systems describing by Hamiltonians similar to (3) can be found in liquid-state NMR [20]. Actually, isotropic $J$-couplings between the nearest neighbour spins are much stronger than those between the remote spins. Therefore, a linear chain of nuclear spins behaves almost as a chain with the nearest-neighbour interactions. Truncation of the isotropic $J$-coupling to a $ZZ$-term can result from large difference between the chemical shifts of the neighbour spins. The domino dynamics mentioned above is observed at the condition $\omega_1 \ll J \ll \omega_0$. Using the unitary transformation,

$$\rho_{rot}(t) = U(t)\rho(t)U^+(t) \qquad (4)$$

with the operator $U(t) = \exp\left(-it\sum_{n=1}^{N}\omega_0 I_n^z\right)$, the evolution equation (2) can be transform to the called "Zeeman interaction representation" [22, 23], i. e. to the frame rotating with the frequency $\omega_0$. The unitary transformation (4) does not change any observable values, such as the individual polarization $P_n(t) = Tr(\rho_{rot}(t)I_n^z) = Tr(\rho(t)I_n^z)$. In the rotating frame at the condition $\omega_0 \gg J \gg \omega_1$, the fast oscillating terms with frequencies $\omega_0$ and $2\omega_0$ can be omitted [22, 23] and the Eq. (2) reduces to the following form

$$i\frac{d\rho_{rot}(t)}{dt} = [H_{rot}, \rho_{rot}(t)] \qquad (5)$$

where the time-independent Hamiltonian

$$H_{rot} = \omega_1 \sum_{n=1}^{N} I_n^x + J \sum_{n=1}^{N-1} I_n^z I_{n+1}^z \qquad (6)$$

describes the 1D Ising model in a transverse magnetic field [21]. Then the density operator, $\rho_{rot}(t)$, develops according to the solution of Eq. (5)

$$\rho_{rot}(t) = e^{-iH_{rot}t}\rho_{rot}(0)e^{iH_{rot}t} \qquad (7)$$

where $\rho_{rot}(0)$ is the initial density matrix.

The model considered by us differs from the usual Ising model which has a magnetic field directed along the $Z$-axis. This last case has no entanglement [24, 25], since the Hamiltonian is diagonal in the standard disentangled basis $\{|00\rangle, |01\rangle, |10\rangle, |11\rangle\}$, where $|0\rangle$ stands for spin down and $|1\rangle$ stands for spin up. However, if the magnetic field is not parallel to the $Z$-axis, it is sufficient to make the eigenstates entangled [25].

### III. Reduced density matrix

In order to quantify the entanglement, the concurrence, $C$, is usually used [26, 27]. For maximally entangled states, the concurrence is $C = 1$ while for separable states $C = 0$. The concurrence between a pair of the spins $m$ and $n$ is expressed by the formula

$$C_{mn}(t) = \max\left\{0, 2\lambda(t) - \sum_{k=1}^{4} \lambda_k(t)\right\}, \tag{8}$$

where

$$\lambda(t) = \max\{\lambda_1(t), \lambda_2(t), \lambda_3(t), \lambda_4(t)\}, \tag{9}$$

and $\lambda_1, \lambda_2, \lambda_3,$ and $\lambda_4$ are the square roots of the eigenvalues of the product

$$R_{mn}(t) = \rho_{mn}(t)\tilde{\rho}_{mn}(t) \tag{10}$$

with

$$\tilde{\rho}_{mn}(t) = Q\rho_{mn}^*(t)Q \tag{11}$$

and

$$Q = \begin{pmatrix} 0 & 0 & 0 & -1 \\ 0 & 0 & 1 & 0 \\ 0 & 1 & 0 & 0 \\ -1 & 0 & 0 & 0 \end{pmatrix} \tag{12}$$

$\rho_{mn}(t)$ is the two-spin density matrix, the so called "reduced density matrix", defined by

$$\rho_{mn}(t) = Tr_{mn}(\rho_{rot}(t)) \tag{13}$$

for $m$-th and $n$-th spins, $Tr_{mn}(...)$ denotes the trace over the degrees of freedom of all spins except for the $m$-th and $n$-th ones, $\rho_{mn}^*(t)$ is the complex conjugation taken in the $I^z$ representation.

The typical initial equilibrium conditions usually used in the NMR experiments suppose that the spin system in a high-temperature approximation is described by the density matrix $\rho(0) = \sum_{n=1}^{N} I_n^z$ (here we used $\hbar = 1$). We shall study the spin system which is initially in a pseudopure state [28, 29]. Conveniently, the quantum algorithms start with a pure ground state where populations of all states except the ground state are equal to zero. The realization of a pure state in a real quantum system, such as a spin system requires extremely low temperatures and very high magnetic fields. To overcome this problem, a so-called "pseudopure" state was introduced [28, 29]. The density matrix of the spin system in this state can be partitioned into two parts. The first part of the matrix is a scaled unit matrix, and the second part corresponds to a pure state. The scaled unit matrix does not contribute to observables and it is not changed by unitary evolution transformations. Therefore, the behaviour of a system in the pseudopure state is exactly the same as it would be in the pure state. We numerically simulated the dynamics of a Ising chain and considering three groups of the initial conditions:

(i) spin system is initially in a pseudopure state with all spins up described by the density matrix

$$\rho(0) = |1\rangle_1 \otimes |1\rangle_2 \otimes ... \otimes |1\rangle_N \ ; \qquad (14)$$

(ii) spin system is initially in a pseudopure state with all spins up except the first spin, which is down; the system is described by the density matrix

$$\rho(0) = |0\rangle_1 \otimes |1\rangle_2 \otimes ... \otimes |1\rangle_N \ ; \qquad (15)$$

(iii) the spin system is initially in a pseudopure state with all spins up except the first two which are entangled; the density matrix is

$$\rho(0) = \rho_{12}(0) \otimes |1\rangle_3 \otimes ... \otimes |1\rangle_N \ . \qquad (16)$$

The method of creating the highly polarized spin states (14) in clusters of coupled spins (the first initial condition) was described previously [28, 29]. It is based on filtering multiple-quantum coherence of the highest order, followed by a time-reversal period and partial saturation. The initial state with all spins up except the first spin which is down (15) can be prepared using partial saturation and applying a selective Gaussian pulse [20]. Below we will discuss the preparation of a partially entangled initial state (16).

The numerical simulation of the concurrence dynamics of the chains up to 7 spins is performed using the developed software based on the MATLAB package. The calculations are carried out in two stages: first the evolution of the density matrix is simulated using solution (7) and then concurrence of two spins is determined using Eqs. (8)-(13).

**IV. Selective initialization of partially entangled states**

Selective entanglement can be formed in an Ising chain where the spins have different resonance frequencies. In the case when the irradiation acts selectively on two spins, *n*-th and *m*-th, the Hamiltonian (6) takes the following form:

$$H_{sel} = \sum_{k=m,n} \omega_1^k I_k^x + J \sum_{k=1}^{N-1} I_k^z I_{k+1}^z . \qquad (17)$$

The results of the individual polarization, $P_n(t)$ and concurrence, $C_{mn}(t)$ simulations as functions of the irradiation time are presented in Figs. 1 and 2. The change of the polarization of these two spins, while the polarization of the others is kept, shows that the selective action is realized. Figure 1 presents the results for the case when the selective irradiation acts on the *n*-th and *n* +1-th spins placed in the middle part of the chain, i. e. 2 ≤ *n* ≤ *N* - 2. The plots in Fig. 1 illustrate also the dependence of concurrence on the strength of the irradiating field. The results are the same for any *n* and for the initial states (14) or (15) and do not depend on the number of spins *N*.

When the pair of spins is at the beginning of the chain (*m* = 1, *n* = 2), the concurrence dynamics depends on the initial conditions (Fig. 2). For the initial state (14) with all spins up, the maximum concurrence between two first spins, $C_{1,2} = 0.67$ and $0.70$, is achieved at the times $t \approx 4.82/J$ and $10.54/J$, respectively (Fig. 2a). For the initial state (15) with the first spin down, the maximum possible concurrence between two first spins, $C_{1,2} = 1$, can be achieved at the time $t \approx 5.30/J$ (Fig. 2b). For this instant, the density matrix of the entangled spins has the form:

$$\rho_{12}(0) = \frac{1}{2}\begin{pmatrix} 0 & 0 & 0 & 0 \\ 0 & 1 & -1 & 0 \\ 0 & -1 & 1 & 0 \\ 0 & 0 & 0 & 0 \end{pmatrix}_{12} \qquad (18)$$

and it is used in the initial conditions (16).

An example of the spin system which can be used for the experimental realization of the selective entanglement is a chain of four $^{13}C$ nuclear spins of fully $^{13}C$-labeled sodium butyrate described in [20]. This chain can be also described by the Hamiltonian (17). All spins have different resonance frequencies which allow performing the selective radiofrequency irradiation. The maximum possible concurrence between the first two spins $C_{1,2}=1$ can be achieved after the irradiation time $t \approx 0.1 s$.

**V. Evolution of the concurrence of spin pairs**

Let us analyze the dynamics of entanglement between pairs of spins in the system consisting of $N$ spins with Hamiltonian (6) and with various initial conditions (14)-(16). We examine the time dependence of the concurrence, $C_{mn}(t)$, between a pair of the spins $m$ and $n$ by letting the spin system evolve under the Hamiltonian (6) with the coupling constant $J=1$ and $\omega_1 = 0.15$. The initial concurrence equals zero for the initial conditions (14) and (15). For the case of partially entangled initial state, only the concurrence between two first spins $C_{1,2}=1$ and all the concurrences between other spins equal zero.

For any initial state, the initial period of evolution is characterized by the change of the concurrence only between the nearest neighbors in the chain (Figs. 3-6). Note that the entanglement between the neighbors appears practically simultaneously in any place of the chain, and the formation time is independent of the chain length (compare Figs. 3, 4 and 6). Then, the entanglement is developed between remote spins. The longer is the distance between spins, the more time the appearance of their entanglement takes. It follows from Figs. 3-5 as well as from Fig. 7. The concurrence between next-nearest neighbors grows more rapidly for the longer chains (Figs. 3b, 4b, 5b for $C_{1,3}$ and Fig.6b for $C_{N-2,N}$). For more remote spins, the time of the appearance of concurrence increases with the chain length (Figs. 3c, 4c, 5c for $C_{1,4}$).

Direct interactions are usually considered as the factor required for the entanglement creation [10, 16]. Here we observe the generation of the entangled states between remote particles which do not interact directly.

The effect of the initial conditions manifests itself in the shape of the concurrence curves and in some difference of times of appearance of the entanglement between remote spins. The main difference of the case with the initial conditions (16) is the decrease of concurrence between two first spins at the beginning of the process (Fig. 5a).

## VI. Conclusion

We have studied numerically the entanglement dynamics in one-dimensional Ising chains with nearest-neighbor interactions under the various initial conditions. It was shown that, in spite of the use of a model with the direct interactions between the nearest neighbors, the entanglement between remote spins can be generated. The process of the appearance and transportation of the entanglement in the chain depends slightly on initial conditions. Unexpected behaviour of the next-nearest-neighbor concurrence was obtained: the concurrences $C_{1,3}$ and $C_{N-2,N}$ grows more rapidly for the longer chains than for the shorter chains.

Figure Captions:

Fig. 1. (Color online) Time dependence of the individual polarization and concurrence for selectively irradiated two spins $n$ and $n+1$ in the middle part of a chain at various irradiation intensities: $\frac{\omega_1}{J}=0.15$ (a); $\frac{\omega_1}{J}=0.25$ (b); $\frac{\omega_1}{J}=0.35$ (c). The initial state is described by Eq. (14). Black solid line – individual polarization for $n$ and $n+1$-ths spins; blue dash-dot – the concurrence between these spins; green dot – individual polarization for either of rest spins in the chain.

Fig. 2. (Color online) Time dependence of the individual polarization and concurrence for selectively irradiated two first spins in a chain: initial state (14) and $\frac{\omega_1}{J}=0.35$ (a); initial state (15) and $\frac{\omega_1}{J}=0.35$ (b). Black solid line – individual polarization of the first spin; red dash – individual polarization of the second spin; blue dash-dot – concurrence between these spins; green dot – individual polarization for either of rest spins in the chain.

Fig. 3. (Color online) Evolution of concurrences in a chain initially prepared in the highly polarized state with all spins up, Eq. (14), for various pairs of spins: nearest spins $C_{1,2}$ (a), next - nearest spins $C_{1,3}$ (b), and next - next nearest $C_{1,4}$ (c). The curves are given for various numbers of spins in the chain: $N = 4$ – black solid lines; $N = 5$ – red dash; $N = 6$ – green dot; $N = 7$ – blue dash-dot.

Fig. 4. (Color online) Evolution of concurrences in a chain with the initial state (15), where the first spin is down and other spins – up, for various pairs of spins: nearest spins $C_{1,2}$ (a), next - nearest spins $C_{1,3}$ (b), and next - next nearest $C_{1,4}$ (c). $N = 4$ – black solid lines; $N = 5$ – red dash; $N = 6$ – green dot; $N = 7$ – blue dash-dot. Note that the first maximum of the concurrence between the next - next nearest $C_{1,4}$ is about 0.003 and is observed at $Jt = 38$.

Fig. 5. (Color online) Evolution of concurrences in a chain initially prepared in a partially

entangled state, Eq. (16), for various pairs of spins: $C_{1,2}$ (a), next - nearest spins $C_{1,3}$ (b), and next - next nearest $C_{1,4}$ (c). $N = 4$ – black solid lines; $N = 5$ – red dash; $N = 6$ – green dot; $N = 7$ – blue dash-dot.

Fig. 6. (Color online) Concurrence between the next to last spin $N-1$ and the last spin $N$, $C_{N-1,N}$, and concurrence between the next-next-to last and the last spins, $C_{N-2,N}$. Initial condition (15) - the first spin down and other up (a) and (b); initial condition (16) – two first spins are entangled (c). $N = 4$ – black solid lines; $N = 5$ – red dash; $N = 6$ – green dot; $N = 7$ – blue dash-dot.

Fig. 7. (Color online) Generation of entanglement between the ends of a spin chain. Concurrence $C_{1,N}$ for: $N = 4$ – black solid line; $N = 5$ – red dash; $N = 6$ – green dot; $N = 7$ – blue dash-dot. The system is initially prepared in the highly polarized state with all spins up, Eq. (14).

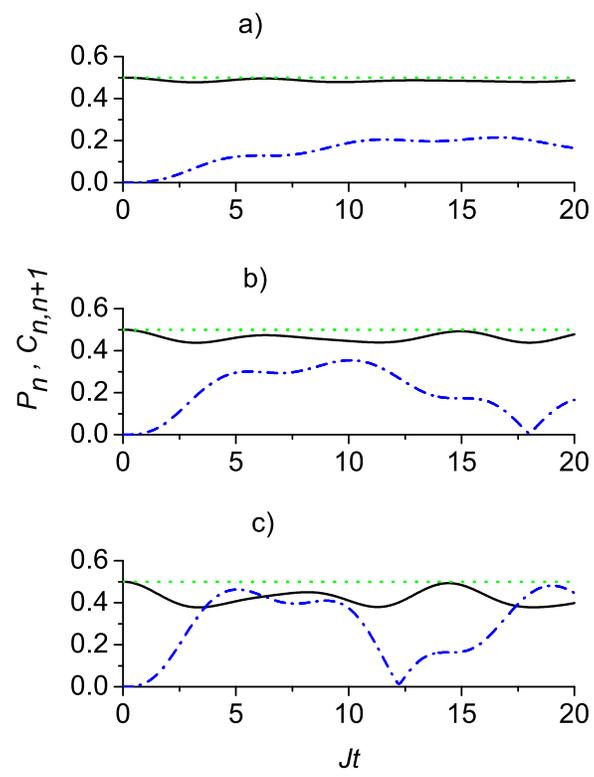

Figure 1

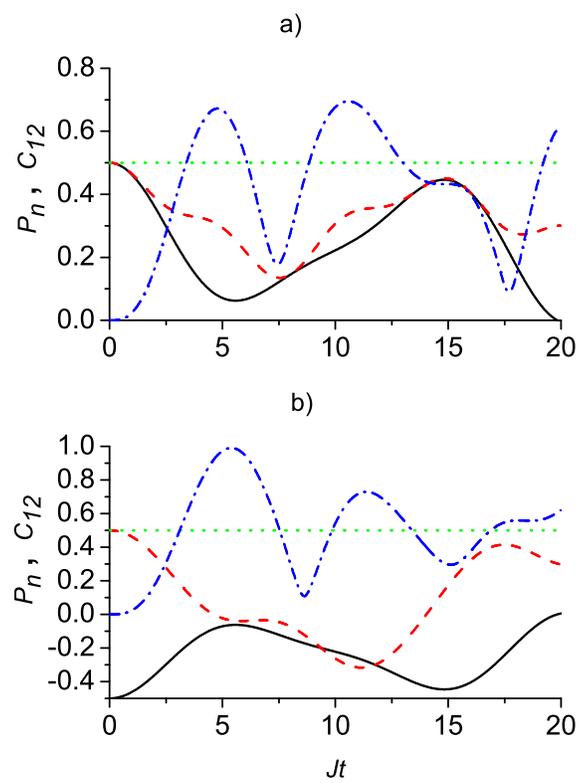

Figure 2

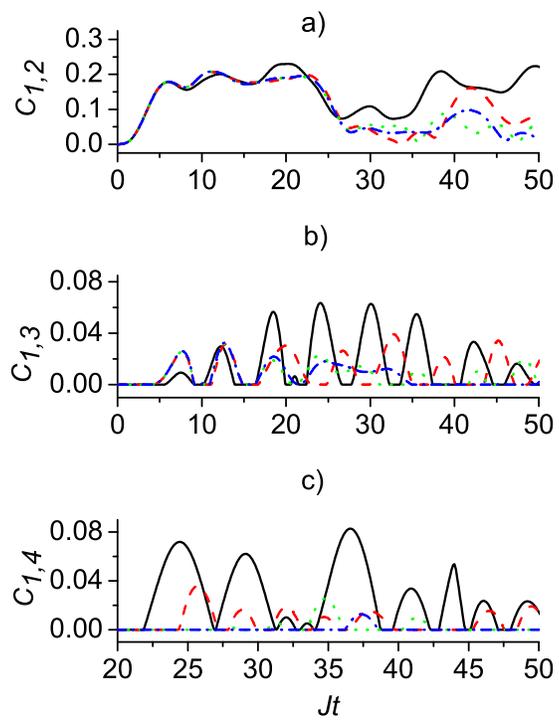

Figure 3

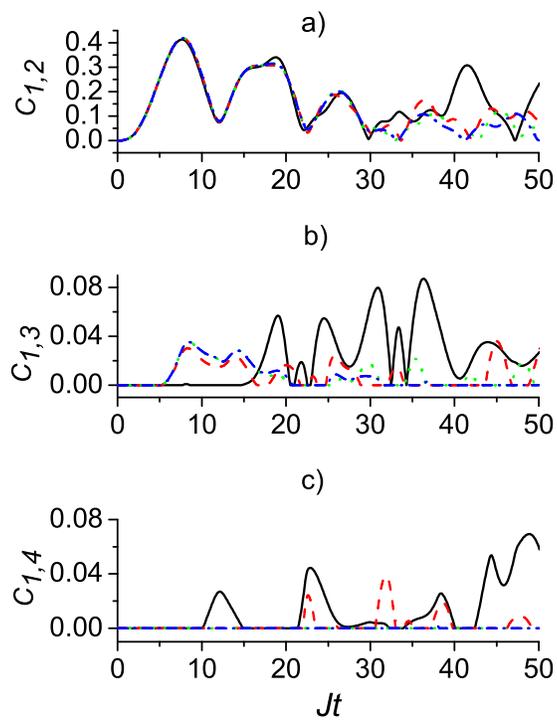

Figure 4

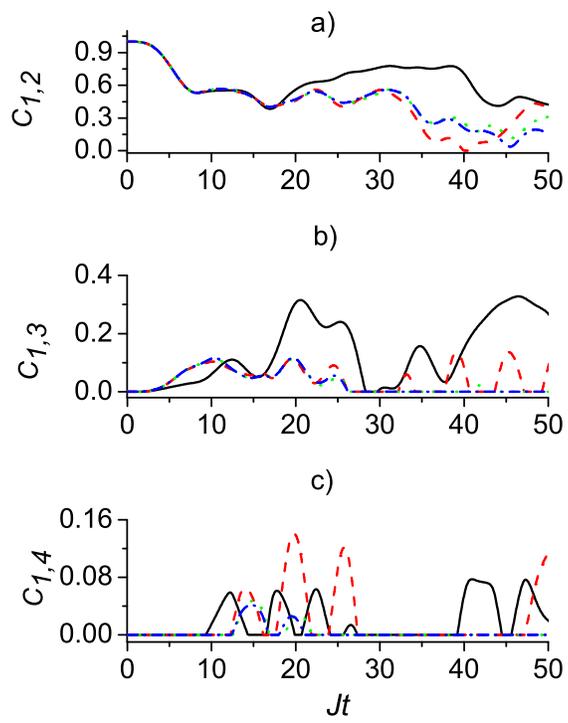

Figure 5

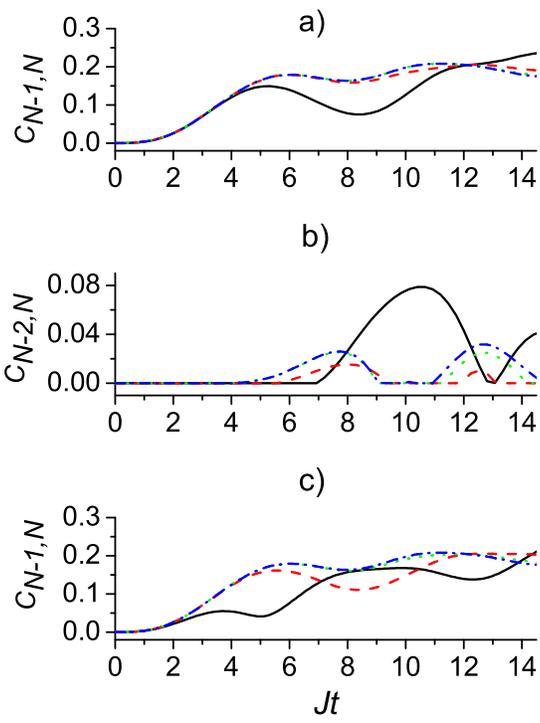

Figure 6

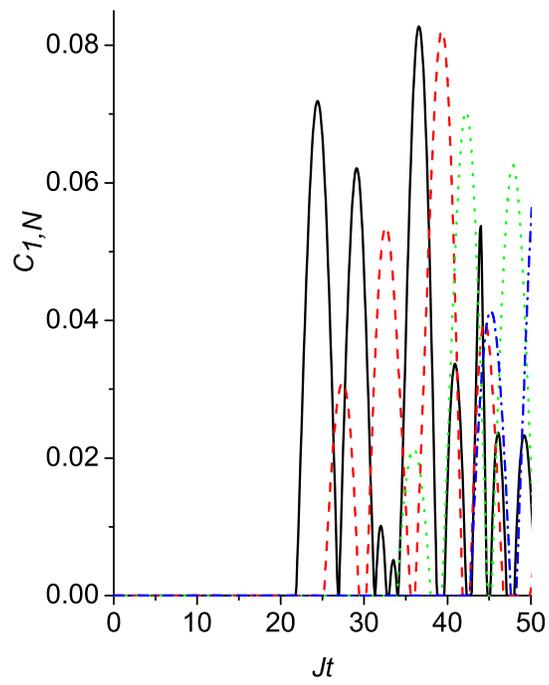

Figure 7